%% file: main.tex
\pdfoutput=1

\documentclass[10pt,twocolumn,letterpaper]{article}

\usepackage[pagenumbers]{cvpr} 

\input{preamble}
\definecolor{cvprblue}{rgb}{0.21,0.49,0.74}
\usepackage[pagebackref,breaklinks,colorlinks,allcolors=cvprblue]{hyperref}

\def\paperID{*****} 
\def\confName{CVPR}
\def\confYear{2026}

\title{Learning to Place Objects with Programs and Iterative Self Training}

\author{Adrian Chang$^{1,2}$\quad Kai Wang$^{2}$\quad Yuanbo Li$^{2}$\quad Manolis Savva$^{3}$\quad Angel X. Chang$^{3}$\quad Daniel Ritchie$^{2}$\\[0.5em]
{\small $^{1}$Vision Systems Inc.}, 
{\small $^{2}$Brown University}, 
{\small $^{3}$Simon Fraser University}
}

\begin{document}
\maketitle
\begin{figure}[ht]
  \centering
  \includegraphics[width=1.0\linewidth]{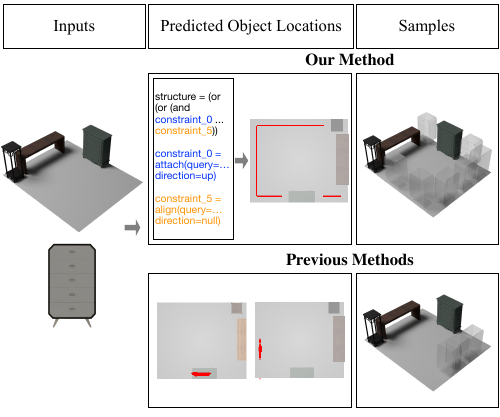}
   \caption{
         Indoor scene object placement is the task of predicting possible placement locations from a given 3D scene and object. We visualize the placement options produced by our and previous methods for the inputs shown on the left. Note how previous methods predict incomplete distributions while ours produces a variety of placement options. 
   }
   \label{fig:teaser}
\end{figure}
\input{sections/0-abstract}
\input{sections/1-introduction}
\input{sections/2-relatedwork}
\input{sections/3-overview}
\input{sections/4-language}
\input{sections/5-program}
\input{sections/6-selftraining}
\input{sections/7-results}
\input{sections/8-conclusion}
\input{sections/9-ethics}

\section*{Acknowledgements}
Thank you Kenny Jones for the early conversations on program induction. Thank you Sheridan Feucht for the discussions and support. Thank you James Tompkin for the helpful feedback in early drafts. Thank you Luca Fonstad and Cal Nightingale for their help on the project. Thank you to everyone who volunteered their time to help with annotation. Funding was provided by NSF award \#1941808. 

{
    \small
    \bibliographystyle{ieeenat_fullname}
    \bibliography{main}
}

\clearpage
\appendix
\input{sections/10-appendix}

\end{document}

%% file: sections/0-abstract.tex
\begin{abstract}
In this work we study indoor scene object placement. Given a 3D indoor scene and an object, the task is to predict placement locations within the scene. Empirical observations of data-driven approaches to the problem show their tendency to miss placement modes. We introduce a system which helps to address this flaw. We design a Domain Specific Language (DSL) that specifies object relational constraints. Upon execution, programs from our language predict possible placements from a partial scene and object. We design a generative model which writes these programs automatically. Available 3D scene datasets do not contain programs to train on, and naively extracted programs only predict the original placement location of scene objects. Training on these programs results in subpar performance so we introduce a new program bootstrapping algorithm that improves our system’s performance compared to the naive approach. To quantify our qualitative observations, we introduce a new evaluation procedure which captures how well a system models per-object location distributions. We ask human annotators to label all the possible places an object can go in a scene and compare this set against locations produced by the system in question. Our system produces per-object location distributions more consistent with human annotators than those produced by existing data-driven approaches and a zero-shot approach using an LLM. While other systems degrade in performance when training data is sparse, our system does not degrade to the same degree. 
\end{abstract}

%% file: sections/1-introduction.tex
\section{Introduction}
\label{sec:intro}

Placing objects in an indoor environment is a natural consequence of spending time in one. A variety of applications rely on predicting possible object locations from a given 3D scene and object. Computer vision and robotics researchers use this information in settings such as scene understanding and robot interaction ~\cite{gadre2022csr, puig2023habitat3, ramrakhya2024seeing}. Users of augmented / virtual reality (AR/VR) and interactive scene generation applications ~\cite{Planner_5D, RoomSketcher_2024} might want to automatically visualize placement options of an object in their indoor space. Predicting possible object locations is also a fundamental subroutine of several scene synthesis systems ~\cite{Paschalidou2021NEURIPS, Ritchie2018FastAF}. 

Data-driven approaches to this problem ~\cite{Paschalidou2021NEURIPS, Ritchie2018FastAF} train on indoor scene datasets to learn placement distributions. Empirical observations of object location distributions predicted by these systems reveal their tendency to be incomplete, i.e they omit many plausible locations. For example, a distribution which only places a bed in the corner of an empty room is incomplete, since one could place it along any of the walls. These prior systems overfit to particular object placements seen during training, which limits their overall usefulness for the above applications. 

Standard solutions to reducing overfitting in these systems are unreliable and come at a cost. Scaling the amount of scene data a system trains on can help it learn a more complete distribution, but this solution is expensive. Neural networks are also known to represent the most common inputs over the rare~\cite{Che2016ModeRG}, so these models might still miss placement modes. Model regularization and stopping training early is another option, but in practice it is hard to balance mode coverage with specificity. 

People use object-to-object and object-to-room relationships to guide object arrangement. Existing neural-network-based systems attempt to encode these rules implicitly, making them hard to control and align with our intuitions. On the other hand, symbolic representations can succinctly represent these rules. Their structured representation makes it easier to incorporate prior knowledge, edit the rules they represent, and ensure desirable characteristics such as completeness upon execution. We hypothesize that learning to produce a symbolic representation, such as programs in a Domain Specific Language (DSL), will result in more complete per-object location distributions. 

We propose a new approach to indoor scene object placement. Rather than tasking a generative model with predicting possible object placements, we instead ask it to predict a program in a DSL. Programs in this DSL are relational layout programs which explicitly represent human activity and inter-object relationships. Given a partial scene and object to place, they produce a binary mask representing all possible object locations. We incorporate this language into a learning-based framework to learn how to automatically produce programs from partial scenes and query objects. Our system helps to address the problem of incomplete next object location distributions that plague previous systems. 

We use a transformer-based generative model to produce programs. Available 3D scene dataset contain no ``ground truth'' location programs which can supervise the model. Programs extracted naively with geometric heuristics only predict the original placement location of scene objects. Training on these programs results in subpar performance, so we introduce an iterative self-training scheme which applies the PLAD framework ~\cite{jones2022PLAD} — prior work in unsupervised visual program inference — to this new domain of object location programs. Our self-training algorithm improves our system’s performance compared to the naive extraction approach. 

The original PLAD method assumes access to ground-truth shapes to learn to predict shape programs. Our setting is much harder because the ground truth location distributions which our programs are trying to match are not available. Our method only has access to single location samples during training. Despite this challenge, our method reproduces a good portion of these placement distributions. For each existing program, our approach proposes new programs, filters out noisy suggestions, and then combines ``good'' candidate programs together. Iteratively repeating this process results in programs that predict a variety of placement locations. 

To quantify the performance of object location prediction methods, we also introduce a new evaluation procedure which captures how well a system models per-object location distributions. We ask human annotators to label all the possible places an object can go in a scene and compare this set against locations produced by the system in question. Our system produces per-object location distributions more consistent with human annotators than those produced by existing data-driven approaches and a zero-shot approach using an LLM. While other systems show consistent degradation in per-object location modeling with less scene data, our system does not degrade in performance to the same degree. 

In summary, our contributions are: 
\begin{itemize}
    \item A new approach for placing objects in indoor scenes where we predict a relational layout program from a given partial scene and object to place, and execute that program to predict possible object placement locations
    \item A new bootstrapped self-training algorithm that adds placement modes to naive single location programs by iteratively proposing, filtering, and then combining programs together
    \item A new evaluation procedure that evaluates a system's ability to model per object location distributions
\end{itemize}

%% file: sections/2-relatedwork.tex
\section{Related Work}
\label{sec:related_work}

\textbf{Indoor scene synthesis and object placement.} Before the existence of large indoor scene datasets~\cite{fu20213d} and 3D deep learning algorithms, researchers positioned objects in scenes with explicit rules such as statistical relationships between objects~\cite{Yu2011MakeIH}, programmatically-defined constraints ~\cite{Yeh2012SynthesizingOW}, design principles~\cite{Merrell2011InteractiveFL}, or heuristics for human activity~\cite{Fu2017AdaptiveSO, Fisher2015ActivitycentricSS}. Similarly, our DSL also encodes object relationships and human activity explicitly.   

Deep learning enabled a variety of approaches for learning scene priors from large datasets. A scene graph is a popular representation with works using a graph neural network~\cite{Wang2019PlanIT, gao2023scenehgn}, recursive neural network~\cite{Li2018GRAINS} or diffusion model~\cite{tang2024diffuscene, sun2025reltriple, Hu24arxiv-MiDiffusion} to learn these priors. Image based approaches operate over the top down view of the scene with CNNs~\cite{Ritchie2018FastAF, Wang2018DeepCP}. Transformer based approaches~\cite{Paschalidou2021NEURIPS, Para2020GenerativeLM, wang2020sceneformer} found success working directly with the 3D bounding box information of objects in the scene. Our system leverages recent advances in 3D deep learning, using the transformer architecture as the backbone for scene generation. Instead of having the network directly predict object placements however, our generative model writes programs defined by our DSL. 

Other works source scene priors beyond information distilled from 3D scene datasets. Zero-shot approaches leverage the latent scene knowledge embedded within Large Language Models (LLMs) to position objects and generate scenes. LayoutGPT ~\cite{feng2023layoutgpt} directly predicts the positions and orientations of objects. Predicting constraints and then solving them for possible placement locations is another popular approach ~\cite{yang2023holodeck, Aguina-KangOpen, sun2024layoutvlm, huang2025fireplace}. We also generate and then solve constraint programs. Our method however relies solely on information within 3D scene datasets. 

\textbf{Visual Program Inference.} Visual Program Inference (VPI) aims to automatically infer programs that explain visual data~\cite{Ritchie2023NeurosymbolicMF}. If the visual data of interest comes with ground truth programs, supervised learning is an obvious option~\cite{Wu_2021_ICCV, fusion_360, xu2022skexgen}. In most domains however, programs for visual data are not readily accessible. Unsupervised learning, and in particular bootstrapping~\cite{dreamcoder, Liang2016NeuralSM}, is one option for extracting and improving programs from visual data. 

Bootstrapping methods search for ``good'' programs, retrain on these programs, and then repeat. PLAD~\cite{jones2022PLAD} groups different bootstrapping methods under a single conceptual framework and applies it to VPI on 3D and 2D shapes. Later works which built on the PLAD framework~\cite{ganeshan2023coref, jones2024VPIEdit} searched for new programs by editing existing ones. SIRI~\cite{ganeshan2023coref} used domain specific operations to edit a subset of programs at a time for distributional stability. Our algorithm is an instance of PLAD, and like SIRI, we use domain-specific editing operations.

%% file: sections/3-overview.tex

\begin{figure}[t]
\includegraphics[width=\linewidth]{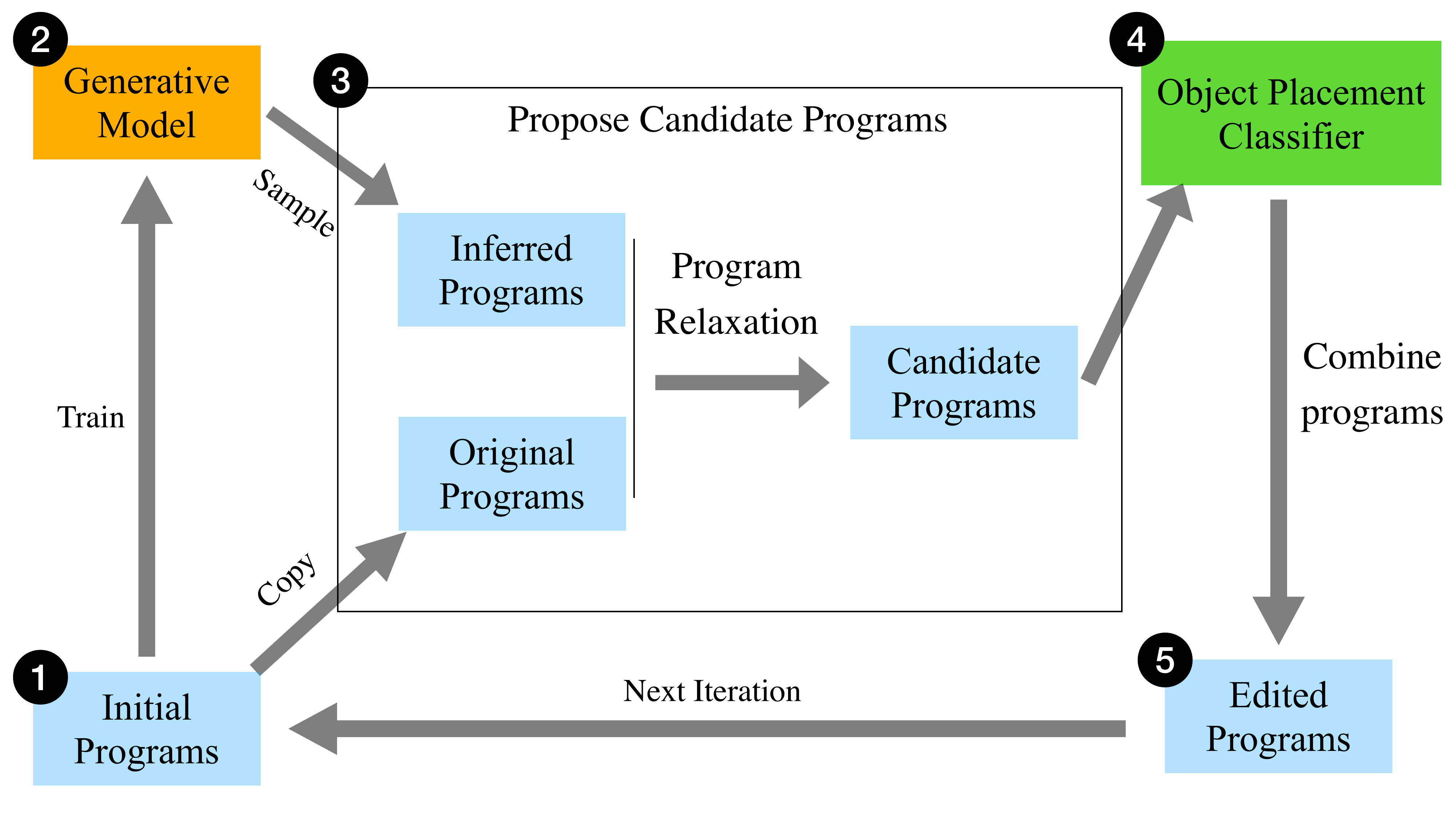}
\vspace{-1.5em}
   \caption{
   Our self-training algorithm discovers programs automatically from scene data. \textbf{(1)} Naively extracted programs serve as initial training data for our \textbf{(2)} model. \textbf{(3)} Deleting constraints from the inferred and original programs produce candidate programs which are then filtered by a \textbf{(4)} classifier. \textbf{(5)} “Good" programs are combined with domain specific operations, and then inserted back into the training set. 
   }
\vspace{-1em}
   \label{fig:self_training_overview}
\end{figure}

\section{Method Overview} 

Figure \ref{fig:teaser} shows our inference pipeline and Figure \ref{fig:self_training_overview} shows how we train our system.
Our full system consists of the following steps:

\textbf{Defining a DSL to describe object placement distributions.} To represent object placement distributions in semantically meaningful terms and align placement rules closer to ones a human might write, we introduce a domain specific language (DSL). Section \ref{sec:language_design} describes this language and its motivations in more depth. 

\textbf{Learning to generate programs.} We design a generative model which writes programs automatically from a partial scene and object to add. Section \ref{sec:generating_programs} describes this model. 

\textbf{Program bootstrapping.} Current indoor scene datasets do not contain programs to train on, so we introduce a program bootstrapping algorithm to discover these programs and boost system performance. Section \ref{sec:program_self_training} describes this algorithm. 

%% file: sections/4-language.tex
\section{Language Design}
\label{sec:language_design}

\begin{figure}[t]
  \centering
  \includegraphics[width=\linewidth]{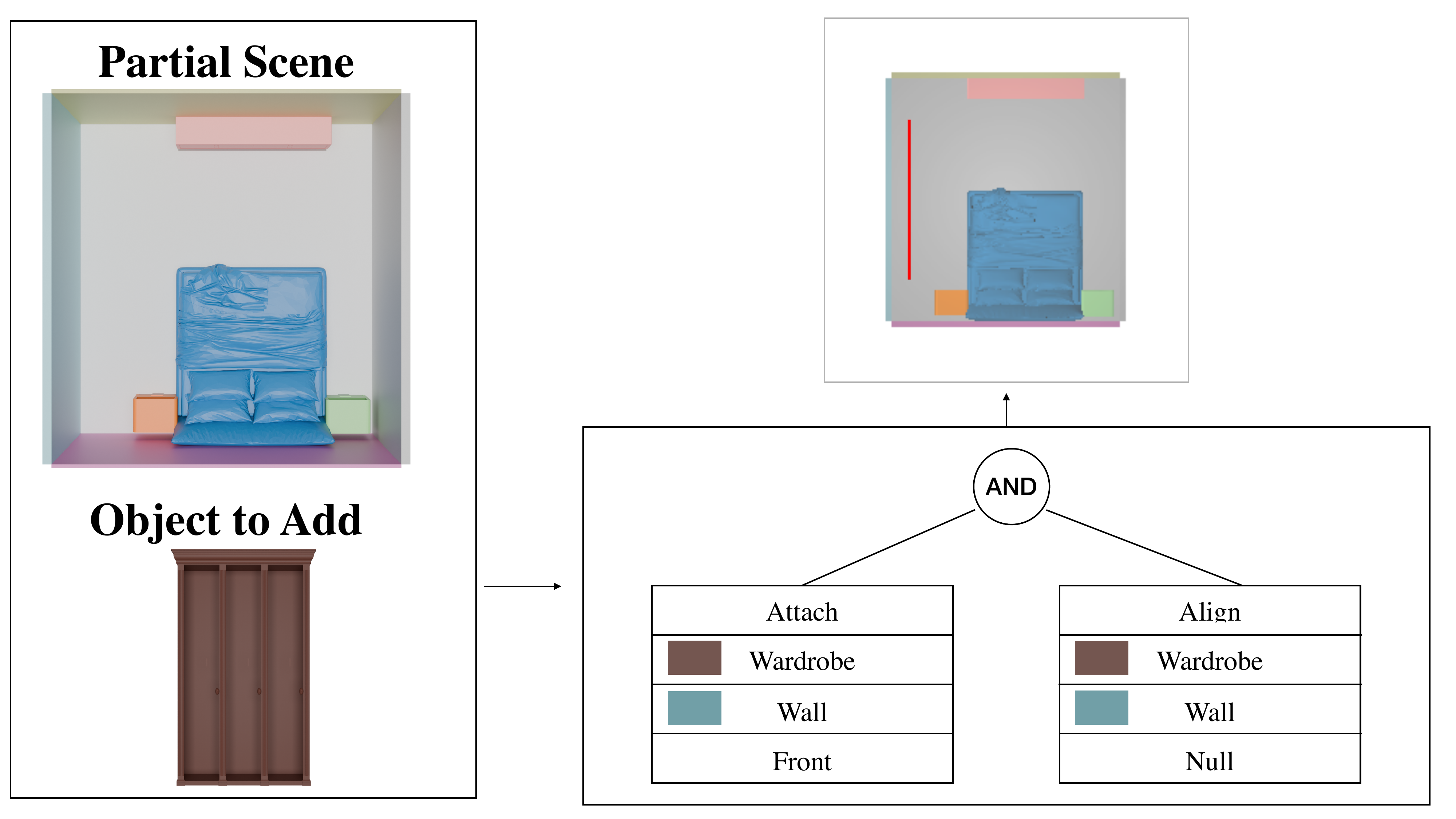}
  \vspace{-2em}
   \caption{Example Program: Given a partial scene and object to add, our DSL program outputs a binary mask representing possible placements of that object. Programs take on the structure of Constructive Solid Geometry (CSG) trees where each leaf node is a constraint that describes object function. Upon execution, these constraints produce binary masks which are combined according to the structure of the tree. }
  \vspace{-1em}
   \label{fig:program_example}
\end{figure}

Our DSL programs take as input a partial scene and the next object to place. They then output a binary mask representing possible centroid locations of the query object. These programs take on the same structure of a Constructive Solid Geometry (CSG) tree, but instead of operating over a continuous 3D space, the programs operate over 2D masks. Leaf nodes of this tree are functional constraints that explicitly represent human activity and inter-object relationships. When executed, these constraints produce binary masks.

Our binary masks are discretized along 3 dimensions: the width of the room, the height of the room, and the possible orientations of the object to place, respectively. This third dimension of the mask is necessary because the validity of an object's centroid position is dependent on its orientation. We represent the orientation of an object as its rotation about the up axis of the room and snap it to one of the cardinal directions (N, E, S, W). Only 6\% of objects in 3D-FRONT ~\cite{fu20213d} deviate more than 10 degrees from one of the cardinal directions so our language can model most object placements. Figure \ref{fig:program_example} shows an example program, its inputs, and its outputs. 

We hypothesize that forcing our system to explain object placements with logical rules and semantically meaningful terms will produce placement rules more consistent with human intuition. CSG is a straightforward choice for translating logical statements into the visual domain and previous work~\cite{CSGNet, ellis_repl} has shown how CSG programs are conducive to to visual program induction.


\subsection{Constraint Specification}
Constraints in our DSL fall under two categories. Location constraints (\textbf{attach} and \textbf{reachable\_by\_arm}) predict an object’s possible centroid locations for every orientation. Orientation constraints (\textbf{align} and \textbf{face}) constrain both location and orientation. 

Five total directions are specified in the language \emph{(Up, Down, Left, Right, Null)}, and all directions are specified within the local coordinate frame of the reference object. To standardize how constraints in the language are represented for neural net processing, all constraints take the same number of arguments. Since orientation constraints do not require a directional argument, we use a special \emph{Null} direction for their direction argument value. 

\begin{itemize}
    \item \textbf{attach(query object, reference object, direction)}: Constrain the possible centroid locations of the query object to be within 15 centimeters of the reference object in the direction specified
    \item \textbf{reachable\_by\_arm(query object reference object, direction)}: Constrain the possible centroid locations of the query object to be between 15-60 centimeters of the reference object in the direction specified. The reference object must also hold humans (i.e. bed, chair). 
    \item \textbf{align(query object, reference object)}: Constrain the possible orientation of the query object such that it points in the same direction as the reference object. 
    \item \textbf{face(query object, reference object)}: Constrain the possible locations of the query object such that it points toward the reference object. Evaluate this for every possible orientation. 
\end{itemize}

Executing a program will execute each constraint in the tree and then combine the masks accordingly. We apply a post-processing step that removes placements of the query object which intersect with other objects in the scene beyond a specified threshold. 

%% file: sections/5-program.tex
\section{Generating programs}
\label{sec:generating_programs}

\begin{figure}[t]
  \centering
  \includegraphics[width=1.0\linewidth]{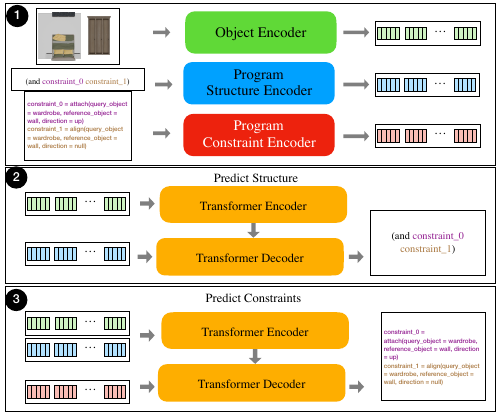}
   \caption{
   Our model predicts DSL programs from a partial scene and object. We formulate this task as a seq-2-seq problem. \textbf{(1)} We vectorize and then embed both the input objects and program. The structure of the program tree and the constraint attributes are embedded as separate sequences. \textbf{(2)} Our first transformer encoder-decoder pair predicts the structure of the program from the input objects. \textbf{(3)} Our second transformer encoder decoder pair predicts the constraint attributes from the object and structure embeddings. 
   }
   \label{fig:generative_model}
\end{figure}

Previous approaches to program synthesis use partial programs to express the high-level structure but leave holes for low level implementation details~\cite{sketch, dreamcoder}. In similar fashion, our model uses one network to predict the topological structure of our program, and then another to fill in the details.  

\subsection{Overview}
We treat program synthesis as a sequence to sequence (seq-2-seq) translation task. The input sequence are the objects in the room and an object to place. The output sequence is the program. We train two transformer~\cite{Vaswani2017AttentionIA} encoder-decoder pairs. The first pair takes in object encodings and outputs the ``structure" of the program. The second pair takes in both the object encodings and program structure and outputs the constraint attributes. Figure \ref{fig:generative_model} shows the network architecture. 

\subsection{Object Encoding}

Objects are represented by their bounding box with attributes category, size, position, orientation, and whether it holds humans $o_i = \{t_i, s_i, p_i, o_i, h_i\}$. The category $t_i$ is an integer id. $s_i, p_i \in \mathbb{R}^2$ (height of objects are not considered and all objects are considered grounded). The orientation $o_i \in \mathbb{R}$ of the object is the rotation of the object about the up vector. Holds\_humans $h_i \in [0, 1]$ is a binary flag of whether the object’s purpose is to hold a human. 

The object encoder encodes object bounding boxes into an embedding vector $\in \mathbb{R}^d$. A learned embedding of the object category is concatenated to the raw values of the other attributes and passed through an MLP. 
Encoding the floor plan, and implicitly the walls, as a single feature vector causes the model to struggle with algebraic quantities such as object to wall distances. This is especially true for floor plans with non-convex geometry. 
Instead, we encode each wall segment as its own object. 

\subsection{Program Encoding and Decoding}
We represent programs as two separate sequences. The first sequence is the program’s tree structure flattened with prefix notation and embedded with per token learnable embeddings. The second sequence are the constraint attributes concatenated following inorder traversal. Each constraint takes the form (constraint type, query object index, reference object index, direction) or $(c_j, q_j, r_j, d_j)$. The constraint type and direction receive per-token learnable embeddings. Tokens which represent the query or reference object index use their respective object embeddings generated by the object encoder. 

For tokens of fixed vocabulary length such as program structure, constraint type, and direction, an MLP head is enough to decode their tokens. The reference object index however has a variable length vocabulary. The number of objects in a scene is varies. To address this problem, we pass each reference object head through an MLP to form a pointer embedding $v_j \in \mathbb{R}^d$~\cite{pointer_networks}. For a matrix of object embeddings $X \in \mathbb{R}^{t \times d}$, where $t$ is the number of objects in the scene, we compute the reference object index $r_j$ as

\begin{equation}
    r_j = \text{argmax}(\text{Softmax}(Xv_j))
\end{equation}

The dot product of the pointer embedding with the object embeddings forms a probability distribution over the objects. The reference object is the object with the highest probability mass. 

\subsection{Alternative Approaches}
Using an LLM to generate DSL programs is another possibility, especially given that our work and many existing zero-shot methods both use constraint-based DSLs ~\cite{yang2023holodeck, Aguina-KangOpen, sun2024layoutvlm, huang2025fireplace}. We find however that few-shot prompting GPT-5\cite{gpt5} to generate programs in our DSL results in hallucinated constraints and only similar performance to our data-driven method, as we will show in Section \ref{sec:evaluation}. Another option could be fine-tuning a LLM at each iteration of our self-training procedure. However, this approach is computationally infeasible. Fine tuning Llama2-7b \cite{touvron2023llama2openfoundation} with LoRA \cite{hu2021loralowrankadaptationlarge} results in a 515 \% increase in training time. 

%% file: sections/6-selftraining.tex
\section{Program Self-Training}
\label{sec:program_self_training}

In this section, we describe our program self-training algorithm which improves next-object distribution locations predicted by our system. An overview of our algorithm is shown in Figure \ref{fig:self_training_overview}. 

Our algorithm falls under the PLAD~\cite{jones2022PLAD} family, a conceptual framework for unsupervised program bootstrapping. These methods iteratively improve a dataset of programs by searching for new and better programs, retraining on them, and then repeating the process. Our algorithm also takes inspiration from Talton et al.~\cite{Talton2012LearningDP}, a work which uses probabilistic context free grammars (PCFGs) to learn a procedural model from a set of examples. Their optimization begins with the “most specific” grammar and converges to a grammar which is not too specific and not too general. 

Our optimization begins with programs that specify a single valid placement. For each object in a scene, we use geometric heuristics to apply every possible constraint to the object so that the extracted program will only place the object where it was originally found. Our iterative self-training pipeline then adds additional placement modes to these ``most restrictive'' programs through a search and filtering process. Similar to Ganeshan et al.~\cite{ganeshan2023coref} we use domain specific operations to edit a subset of programs per iteration of self training. This is both for computational feasibility and distributional stability between iterations. 

\subsection{Candidate Program Generation} 
For a given partial scene and object to add, we search for new programs by sampling the generative model described in Section \ref{sec:generating_programs} and then relaxing both the inferred and original program. Program relaxation involves randomly removing constraints from the program tree to produce a new program. This can help generalize the overly restrictive programs produced by our initial naive approach. For example, an object found in the corner of a room might initially be constrained to rest against both walls. Removing one of these constraints would allow for more general placements along one of the walls. Programs which predict no valid placements or a placement anywhere in the scene are removed from consideration. 

Each candidate program predicts only one valid orientation. Programs whose predictions cover multiple orientations are split into subtrees that each predict only one. This constraint reduces the number of repeated subtrees present in the final programs and improves quantitative performance of this algorithm. It is worth noting that this constraint does not restrict the possible orientations the final program can represent. Our later program combination step (Section~\ref{sec:combprog}) produces programs which predict valid placements for multiple different object orientations.

\subsection{Object Placement Classifier} 
Relaxed programs can produce invalid placements. To prevent these programs from entering the training set, we need to filter out bad programs automatically. Hard positives and negatives for programs are hard to generate, so we train a real fake classifier to predict how in or out of distribution an object placement is. Positive examples are generated by randomly subsampling scenes. Negative examples come from randomly perturbing the rotation and location of a single object. This process can generate false negatives (e.g. perturbing a wardrobe such that it rests against the same wall), but in practice it is sufficient for learning a useful decision boundary. False negatives are rare compared to true negatives. We sample a program's predicted mask multiple times, insert the object in question at those sampled locations, and then compute the object placement classifier's real probability for each insertion. The program's final score is the average of these probabilities. 

Our classifier architecture uses the object encoder described in Section \ref{sec:generating_programs} with an additional learned vector embedding added to the query object vector encoding. 


\subsection{Combining Programs}
\label{sec:combprog}
By this stage in the pipeline we have, for a partial scene and object to place, multiple programs whose placement distributions have been scored by a scene classifier. Each program is also guaranteed to predict a placement distribution with only one possible orientation. We combine programs whose score is above a preset threshold to produce a new final program on which our generative model will be retrained.

Recall that our programs takes on the structure of a CSG tree. If two candidate programs predict two different placement modes in a scene, we can combine them into a single program that predicts both modes concurrently. We do this by creating a new tree with an \textbf{or} node as its root; the two programs are its children. This process is repeatable for any number of candidate programs, as the newly combined program is now considered a single entity which we can combine with another program. For every possible orientation our language can represent, we choose the candidate program with the largest mask. These programs are combined together to produce the final program. 

%% file: sections/7-results.tex
\begin{figure*}[ht]
  \centering
  \includegraphics[width=\linewidth]{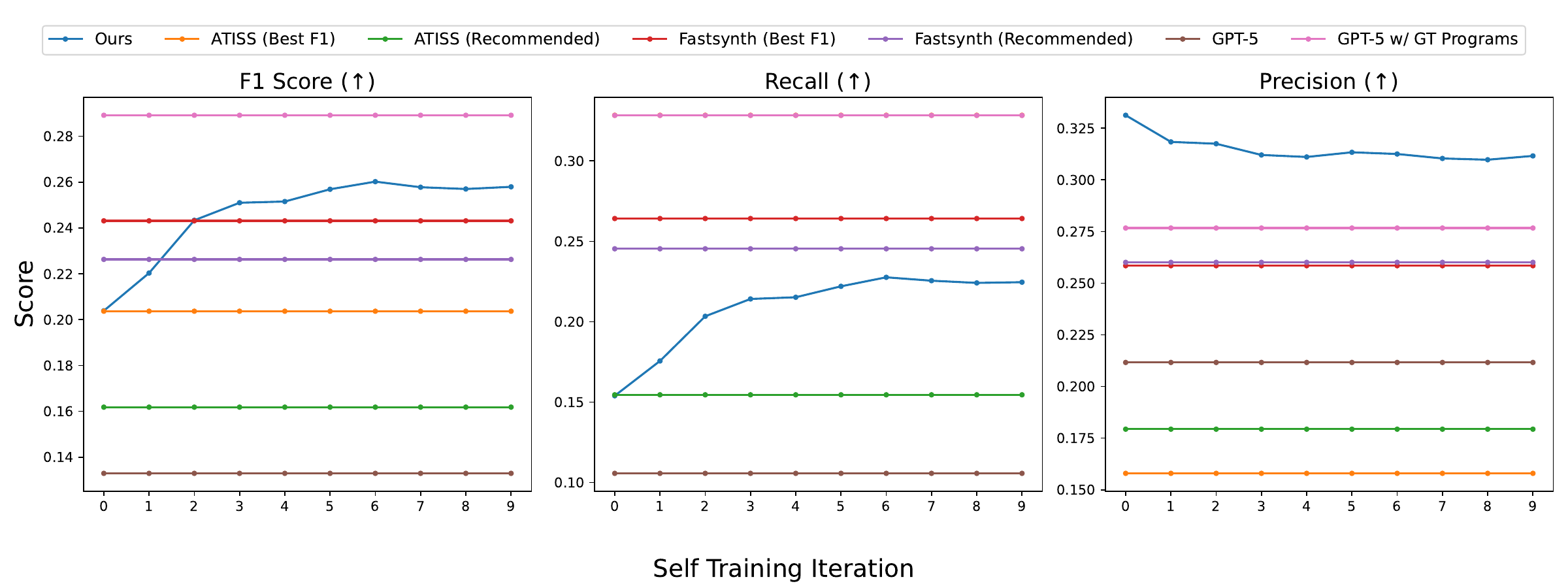}
   \caption{
   We show the F1 score, precision, and recall of per-object location distributions compared against human annotated masks. The x axis is each iteration of our program bootstrapping algorithm. Baseline methods are flat because they do not use self training. 
   Methods denoted with "recommended" use the recommended training time and best performing threshold value. Methods denoted with "best f1" use the settings which maximize F1 score. 
   }
   \label{fig:self_training_metrics}
\end{figure*}

\begin{figure*}[ht]
  \centering
   \includegraphics[width=1.0\linewidth]{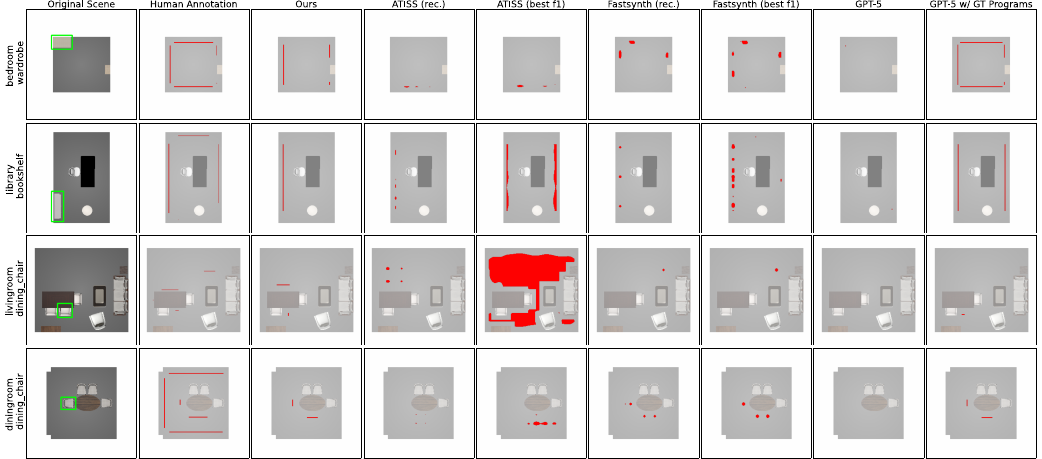}

   \caption{
   Visualization of annotated masks alongside the location distributions predicted by our method, ATISS, Fastsynth, and GPT-5. Green boxes indicate the original placement of the object in the scene. Masks from our method come from the last iteration of self training. The "recommended" set of masks for Fastsynth and ATISS use the recommended training time and best performing threshold. The "best f1" set use the settings which maximize F1 scores. Masks for GPT-5 are produced by few-shot prompting the model to generate our DSL programs. 
   }
   
   \vspace{-0.5em}
   \label{fig:mask_examples}
\end{figure*}

\begin{figure*}
    \centering
    \includegraphics[width=0.9\linewidth]{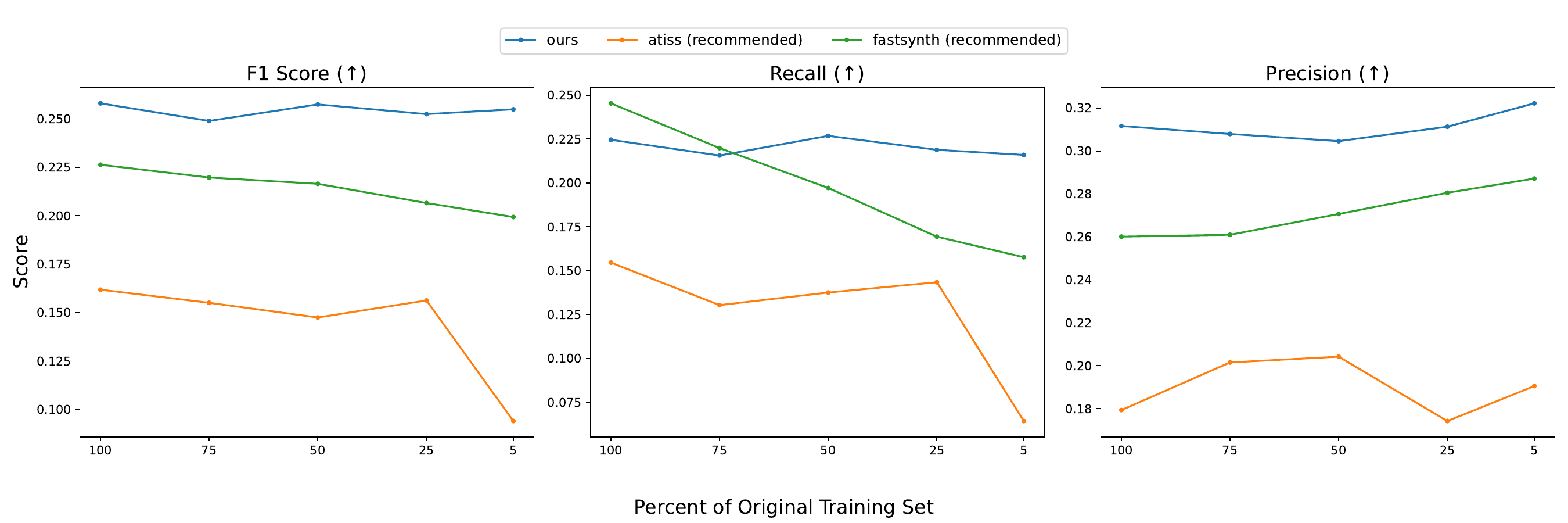}
    \caption{
    Our system maintains performance with as little as 5\% of the original training data. Other systems degrade in the consistency of their predicted per-object location distribution. With less training data recall decreases and precision increases for baseline methods as they overfit to placement locations. 
    }
    \label{fig:data_sparsity_mask_metrics}
\end{figure*}

\begin{figure*}[ht]
  \centering
  \includegraphics[width=0.9\linewidth]{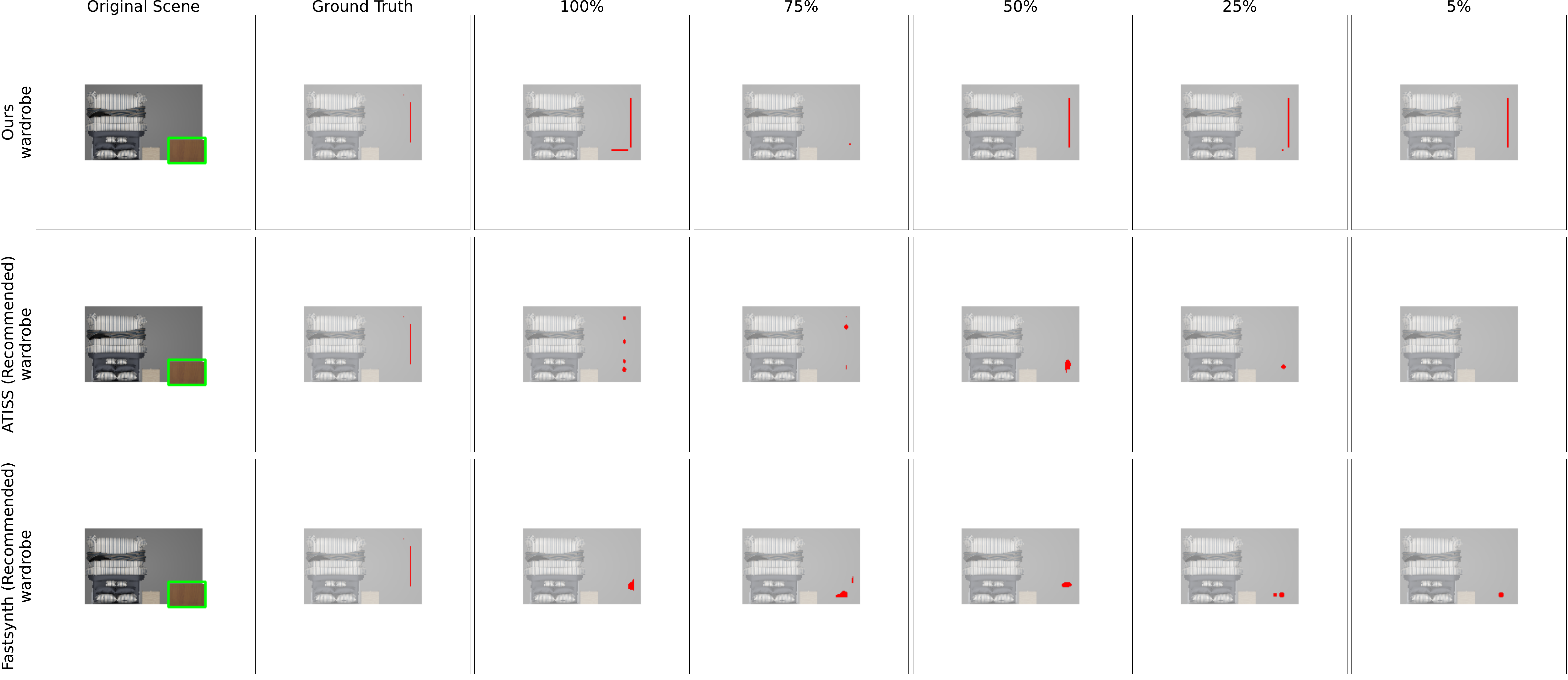}
   \caption{
   Visualizations of how less training data affects the predicted location distributions of our method, ATISS, and Fastsynth. As data becomes sparse, baseline methods collapse to the placement locations seen during training. Our method predicts a variety of placement locations with as little as 5\% of the original training data. It drops only one placement mode in these examples. 
   }
   \label{fig:data_sparsity_mask_examples}
\end{figure*}

\section{Evaluation}
\label{sec:evaluation}
We demonstrate how our system produces per object location distributions that are both more complete and more accurate than than previous methods. We also show our system's superior performance in modeling this location distribution when data is sparse.

Our data-driven baselines are Fastsynth~\cite{Ritchie2018FastAF} and ATISS~\cite{Paschalidou2021NEURIPS} trained on the 3D-FRONT~\cite{fu20213d} dataset. An LLM-based method for object location suggestion does not exist, so we few-shot prompt GPT-5-thinking with chain-of-thought ~\cite{gpt5}. The closest systems do full scene synthesis ~\cite{yang2023holodeck, Aguina-KangOpen, sun2024layoutvlm, huang2025fireplace}, but none have submodules which output location distributions for each object, so we cannot compare against them. We also justify our self training algorithm by evaluating our system without it. 

For Fastsynth and ATISS, we train individual models for four scene types: bedrooms, libraries, living rooms, and dining rooms. We evaluate two versions of each model. One version uses the training time recommended by the original authors, and the other uses the training epoch which maximizes performance on our proposed location distribution metric. 

Our other baseline is GPT-5-thinking few-shot prompted with chain-of-thought to generate our DSL from a partial scene and object. Scene geometry, furniture in the scene, and the query object are converted to a text representation by listing each object and their attributes. DSL programs are converted to text by writing their tree structure with prefix notation and listing the constraint attributes of each leaf node. 

We evaluate two approaches of sourcing which programs to provide as in-context examples in the prompt. Our method receives programs from a heuristically extracted dataset, so the first approach gives GPT-5 the same kind of program data. The example programs in these prompts place objects of the same category as the target object. These programs are flawed however as they are single placement programs that do not benefit from our self-training approach. To measure the upper end of GPT-5 performance, we give the model access to ground truth programs by hand annotating 5 programs for each scene type. GPT-5 performance is expected to improve with access to ground truth. 

\subsection{Comparing Predicted Location Distribution to Human Annotation}
We evaluate a system's ability to model per object location distributions by measuring the Precision, Recall, and F1 Score between a location mask of possible centroid locations predicted by the method in question and a ground truth-mask that comes from human annotators.  

We generated 100 partial scenes and objects for each scene type. Following previous work ~\cite{Ritchie2018FastAF} these partial scenes and objects follow placement order based an object's size and frequency within the dataset. Since our programs output four masks (each representing a different orientation of the object), we collapse them into a single mask for comparison to other methods. For methods that output continuous values, such as ATISS and FastSynth, we generate masks by binarizing the output values with the threshold value that maximizes the F1 score on our validation set. Masks are slightly dilated before comparison to eliminate sensitivity to small discrepancies. 

We show performance of our system over the course of self training as well as performance of the baseline methods in Figure \ref{fig:self_training_metrics}. Our bootstrapping method adds placements modes, increasing recall without sacrificing precision. Baseline methods with high recall have correspondingly low precision because the threshold and training epoch which maximize F1 scores produces distributions that can cover many modes, but are fuzzy and imprecise. Qualitative examples in Figure \ref{fig:mask_examples} show how these distributions predict erroneous locations which often cover a majority of the room layout. Recall drops for these methods when they use the recommended settings because longer training times result in memorized object locations. Note how the data-driven baselines in Figure \ref{fig:mask_examples} only predict portions of the annotated placement rules, often collapsing to single locations. Our method by contrast predicts a variety of possible placements. 

We obtain similar performance to GPT-5 despite having an orders of magnitude smaller model and no ground truth programs to train on. The exact size of GPT-5 is not publicly available, but close open source equivalents ~\cite{llama3} contain hundreds of billions of parameters. Our model has only 2 million. GPT-5 without access to ground truth programs underperforms our method and tends to generate syntactically-incorrect programs or programs with hallucinated constraints (e.g. "reachable\_by\_leg", "center", "front", "horizontal"). When syntactically valid programs are executed, they often result in no valid placements, as shown in Figure \ref{fig:mask_examples}. GPT-5 with access to ground truth programs achieves similar performance to our method, which requires no annotation. Our method can miss placement modes, but our results demonstrate significant improvement in modeling object location distributions when only given access to single location samples.   

\subsection{Measuring Performance With Less Scene Data}
\label{sec:subsection_data_sparsity}

 Our system maintains consistent performance when there are fewer data samples to train on. Previous systems degrade in quality because they require many example placements to learn complete distributions. Our system's program bootstrapping algorithm can generalize sparse samples to more general placement rules. We show with quantitative metrics in Figure \ref{fig:data_sparsity_mask_metrics} and qualitative examples in Figure \ref{fig:data_sparsity_mask_examples} the effect that fewer training examples have on the predicted location distribution. As data becomes sparse, baseline methods collapse to the placement locations seen during training. Our method predicts a variety of placement location and only ever drops one placement mode in the shown examples. 

%% file: sections/8-conclusion.tex
\section{Conclusion}

In this work, we study indoor scene object placement. Our approach uses programs and iterative self training to help address the issue of incomplete location distributions. Our evaluation procedure is the first of its kind, and quantifies the performance of object location prediction methods. 


%% file: sections/9-ethics.tex
\section{Ethics Statement}

Scenes used to train and evaluate our method come from a primarily western design canon and represent only a subsection of the indoor spaces people inhabit. We also design our language with the assumptions that the inhabitants are able bodied. Indoor scenes generated by our method are thus potentially biased against underrepresented groups.

%% file: sections/10-appendix.tex
\begin{figure*}[ht]
  \centering
   \includegraphics[width=1.0\linewidth]{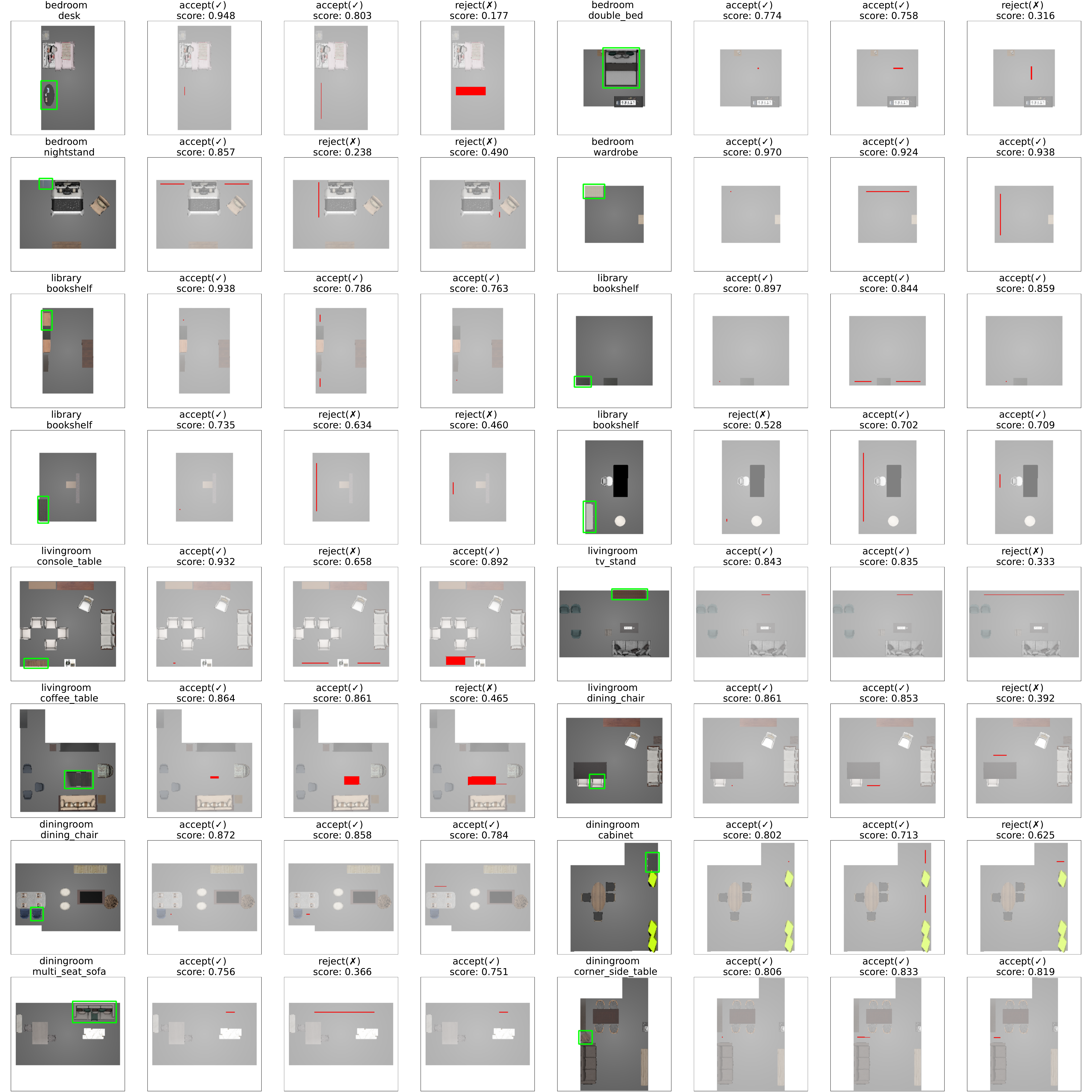}
   \caption{
   Visualization of masks predicted by candidate programs proposed by our self training algorithm alongside their classifier scores. Green boxes indicate the original placement of the object in the scene. All programs come from the first iteration of self training. Program scores are evaluated by sampling the program's predicted mask multiple times, inserting the object in question at those sampled locations, and then averaging the classifier's real probability at each insertion. We use a threshold of 0.7 to accept or reject programs. Programs rarely become overgeneric. The most common failure mode is rejecting programs that predict valid placements.
   }
   \vspace{-0.5em}
   \label{fig:self_training_examples}
\end{figure*}

\section{Scene Synthesis}
\begin{table*}[t!]
\centering
\small
\caption{We report FID, Category KL Divergence, and Scene Classifier Accuracy of generated scenes. Methods denoted with ``best F1'' use the training epoch with the highest F1 score on the location distribution metrics. Methods denoted "recommended" use the recommended training time.  Empty rows mean that the recommended training settings also produce the best F1 scores. Our method generates scenes of comparable quality to previous systems. Although our self training algorithm improves our system's ability to model per object location distributions, they do not significantly hurt or improve its scene synthesis capabilities. ATISS performance degrades when using the epochs with the highest F1 scores because they are often early epochs that have not yet memorized object placements. Fastsynth does not suffer from the same phenomenon. }
\vspace{-0.5em}
\setlength{\tabcolsep}{1.2pt} 
\renewcommand{\arraystretch}{1.2} 
\begin{tabular}{@{}l|ccc|ccc|ccc|ccc|ccc@{}}
\toprule
& 
\multicolumn{3}{c}{\textbf{All}} & 
\multicolumn{3}{c}{\textbf{Bedroom}} & 
\multicolumn{3}{c}{\textbf{Dining room}} & 
\multicolumn{3}{c}{\textbf{Living room}} & 
\multicolumn{3}{c}{\textbf{Library}} \\ 

\cmidrule(lr){2-4} 
\cmidrule(lr){5-7} 
\cmidrule(lr){8-10} 
\cmidrule(lr){11-13}
\cmidrule(lr){14-16}

 & FID $\downarrow$ & CKL $\downarrow$ & SCA \% 
 & FID $\downarrow$ & CKL $\downarrow$ & SCA \% 
 & FID $\downarrow$ & CKL $\downarrow$ & SCA \% 
 & FID $\downarrow$ & CKL $\downarrow$ & SCA \% 
 & FID $\downarrow$ & CKL $\downarrow$ & SCA \% \\
\midrule
\midrule
ATISS (recommended) & 55.245 & 0.0265 & \textbf{76.62} & \textbf{20.148} & \textbf{0.0067} & \textbf{60.89} & 96.063 & 0.0279 & 83.65 & 68.269 & 0.0356 & 80.60 & 36.499 & 0.0359 & \textbf{81.34} \\
ATISS (best F1) & 67.534 & 0.0214 & 82.87 & 20.748 & 0.0068 & 65.44 & 103.062 & 0.0216 & 89.95 & 75.369 & 0.0408 & 86.89 & 104.178 & 0.0252 & 92.71 \\
Fastsynth (recommended) & 60.517 & 0.20434 & 89.80 & 29.617 & 0.220 & 86.38 & 100.976 & 0.1923 & 90.798 & 68.845 & 0.1870 & 91.865 & 42.631 & 0.2180 & 90.163 \\
Fastsynth (best F1) & 61.02 & 0.2043 & 90.671 & - & - & - & 102.311 & 0.199 & 93.00 & - & - & - & 43.308 & 0.2112 & 91.44 \\
Ours (no self training) & 54.456 & 0.02482 & 77.69 & 24.170 & 0.0217 & 71.32 & 91.579 & \textbf{0.0251} & \textbf{80.38} & 64.623 & 0.0321 & 77.67 & 37.453 & \textbf{0.0204} & 81.38 \\ 
Ours & 53.823 & 0.0243 & 78.31 & 22.670 & 0.01133 & 72.52 & 90.665 & 0.0251 & 83.36 & 64.040 & 0.0229 & \textbf{75.76} & 37.916 & 0.0378 & 81.59 \\
\bottomrule
\end{tabular}
\vspace{-0.5em}
\label{tab:scene_synth_metrics}
\end{table*}

\begin{figure*}[ht]
  \centering
   \includegraphics[width=\linewidth]{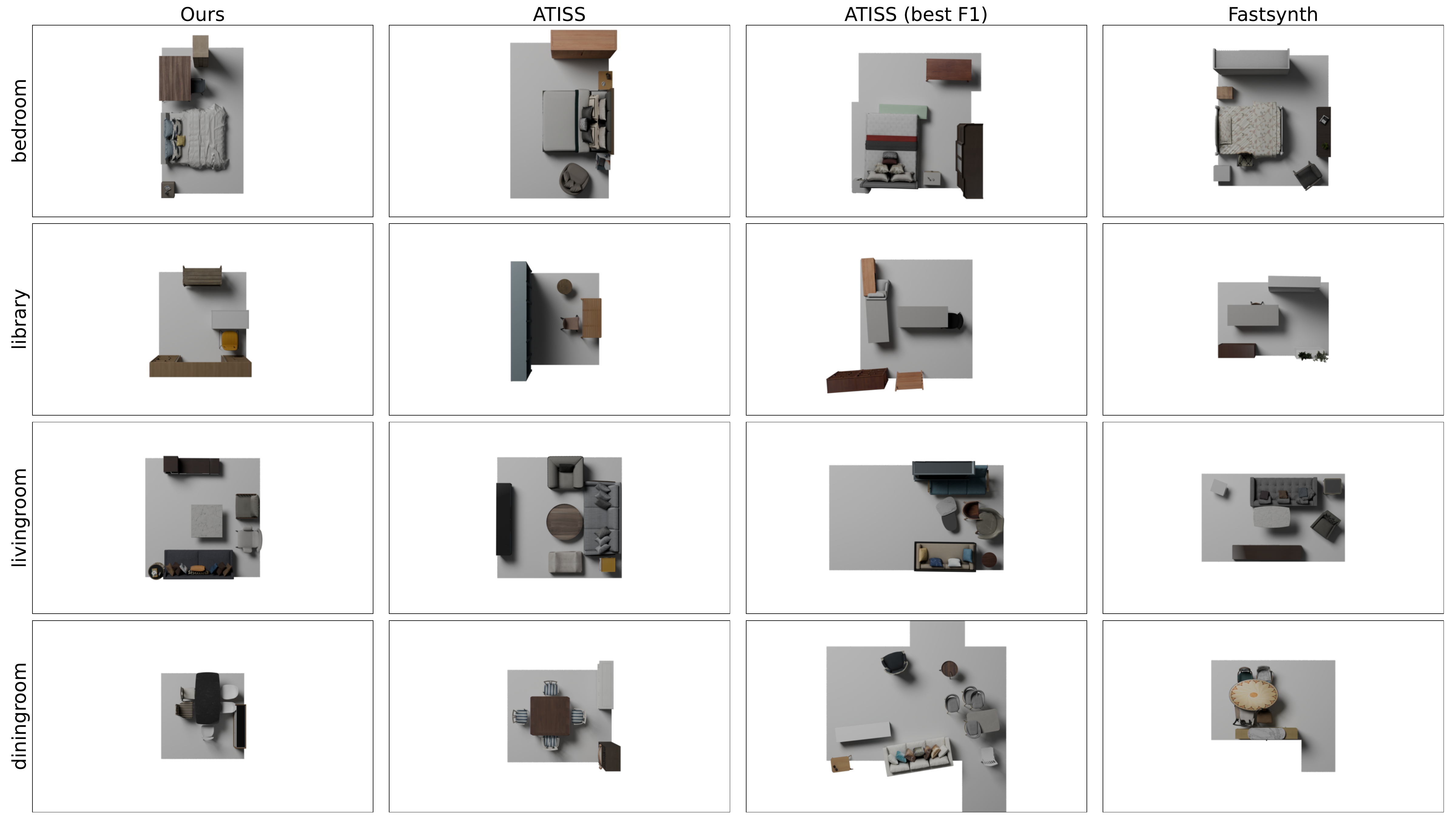}

   \caption{
   Examples of scenes generated by ATISS, Fastsynth, and our method. Our method is capable of generating scenes of comparable quality to previous methods. Note the degradation in final scene quality when ATISS uses the training settings which produce the highest F1 score. 
   }
   \label{fig:scene_examples}
\end{figure*}

\begin{figure*}[ht]
  \centering
   \includegraphics[width=\linewidth]{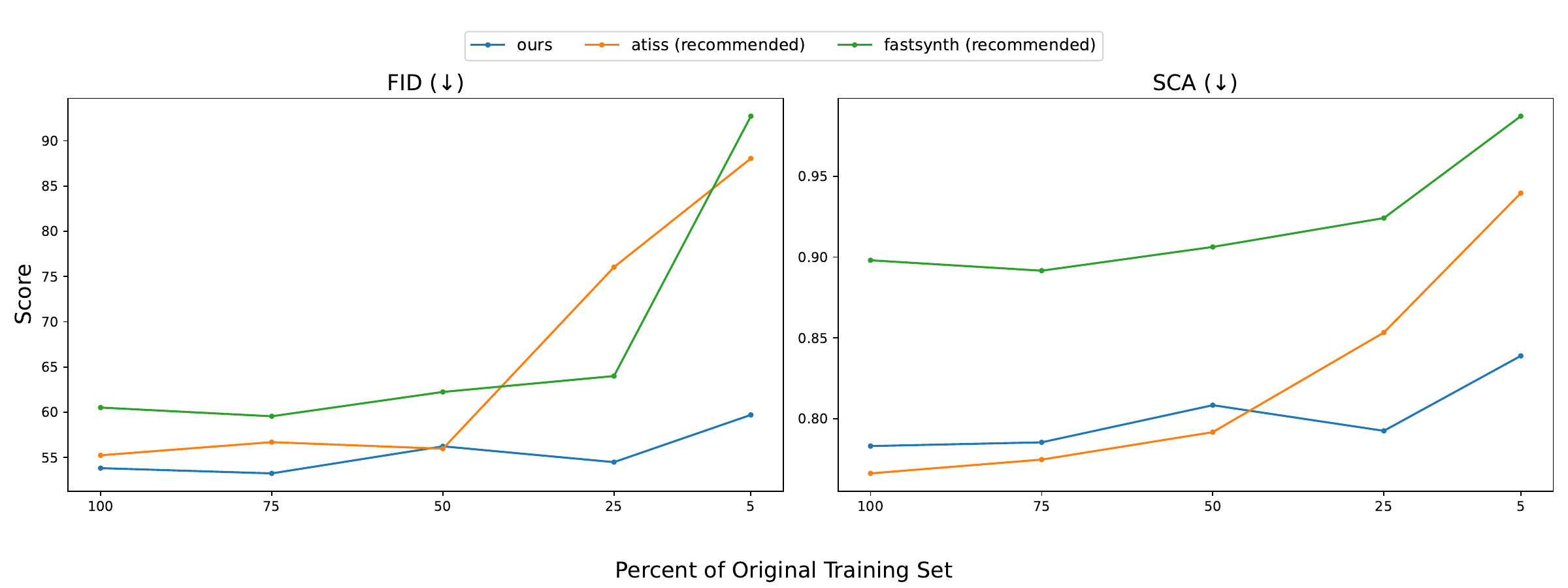}

   \caption{
   Our system maintains scene synthesis performance even as the number of training examples drops to as little as 5\% of the original training set. 
   }
   \label{fig:data_sparsity_scene_synth_metrics}
\end{figure*}

We evaluate our system's ability to perform scene generation from a given floor plan. Our programs do not automatically determine the category and size of the next object to place. We use the category prediction module from Ritchie et al.~\cite{Ritchie2018FastAF} to predict the object category of the next object to place and randomly sample category dimensions from the dataset of objects. 

In accordance with previous work we report the FID scores \cite{fid} of top down orthographic renderings of final scenes, the object category KL divergence between a set of generated scenes and an eval set from 3DFRONT, alongside the real fake object placement classifier accuracy on the two scenes. Object models closest to the bounding box dimensions of each object are chosen from 3D-FRONT~\cite{fu20213d}. Images for FID come from orthographic renderings of generated scenes where each object model receives per-class coloring.  

We finetune a pre-trained AlexNet~\cite{alexnet} to classify these orthographic renderings and report its classification accuracy on a held out test set. 50\% scene classification accuracy indicates that the classifier cannot differentiate between generated and ground truth scenes, so closer to 50\% is better. Our quantitative results are shown in Table \ref{tab:scene_synth_metrics} and qualitative shown in Figure \ref{fig:scene_examples}. 

Our method generates scenes of comparable quality to previous systems. Although our self training algorithm improves our system's ability to model per object location distributions, they do not significantly hurt or improve its scene synthesis capabilities. ATISS performance degrades when using the epochs with the highest F1 scores because they are often early epochs that have not yet memorized object placements. Fastsynth does not suffer from the same phenomenon. 

Section  \ref{sec:subsection_data_sparsity} demonstrated how our system maintains its ability to model per object location distributions while other methods degrade when there are less data samples to train on. We also show this to be true for scene synthesis and graph FID and SCA scores with respect to the number of data samples trained on. Figure \ref{fig:data_sparsity_scene_synth_metrics} shows these results. 

\begin{figure*}[t]
  \centering
  \includegraphics[width=\linewidth]{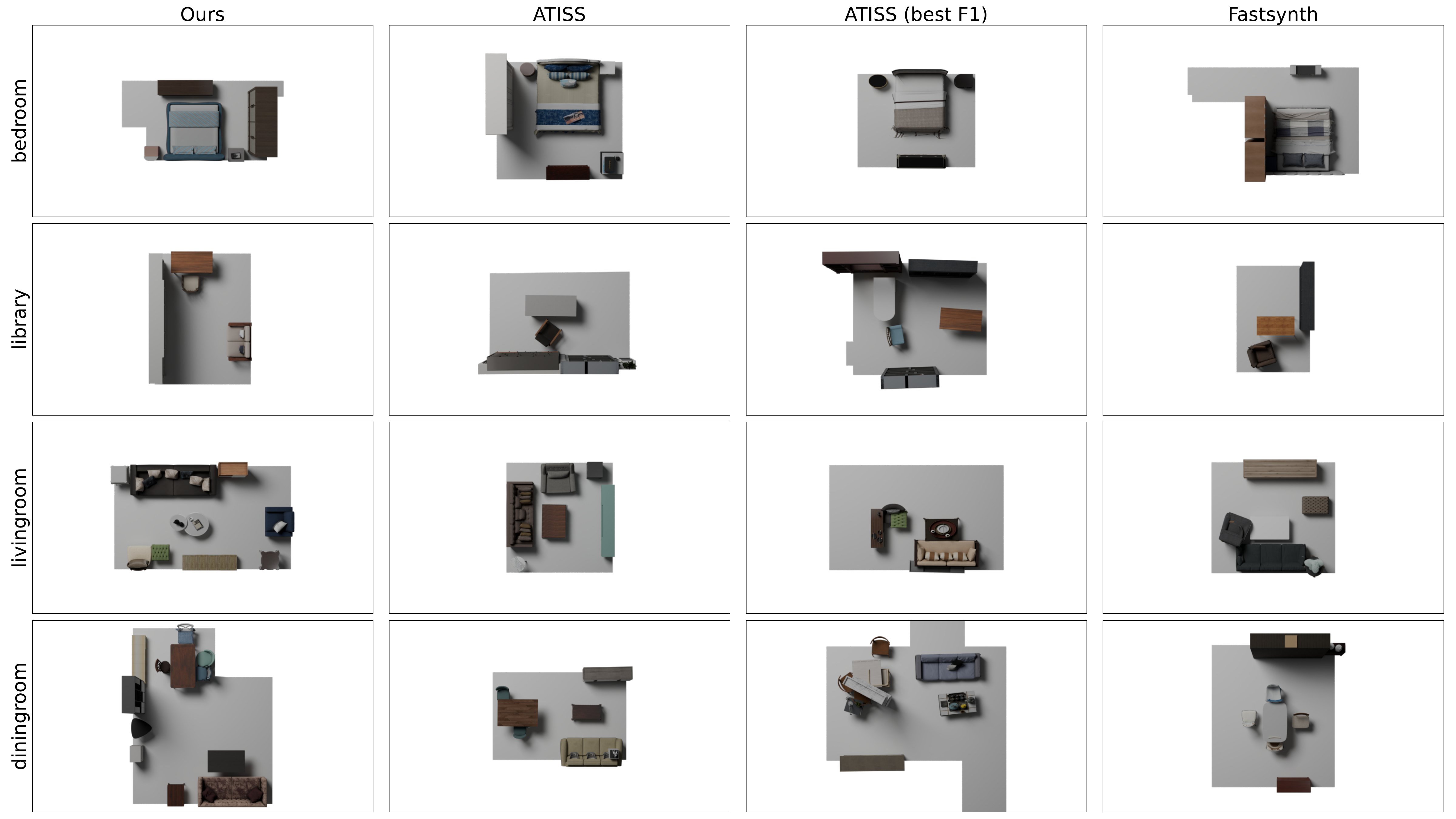}

   \caption{
   More examples of scenes generated by our and baseline methods
   }
   \label{fig:scene_diagram_1}
\end{figure*}

\begin{figure*}[t]
  \centering
  \includegraphics[width=\linewidth]{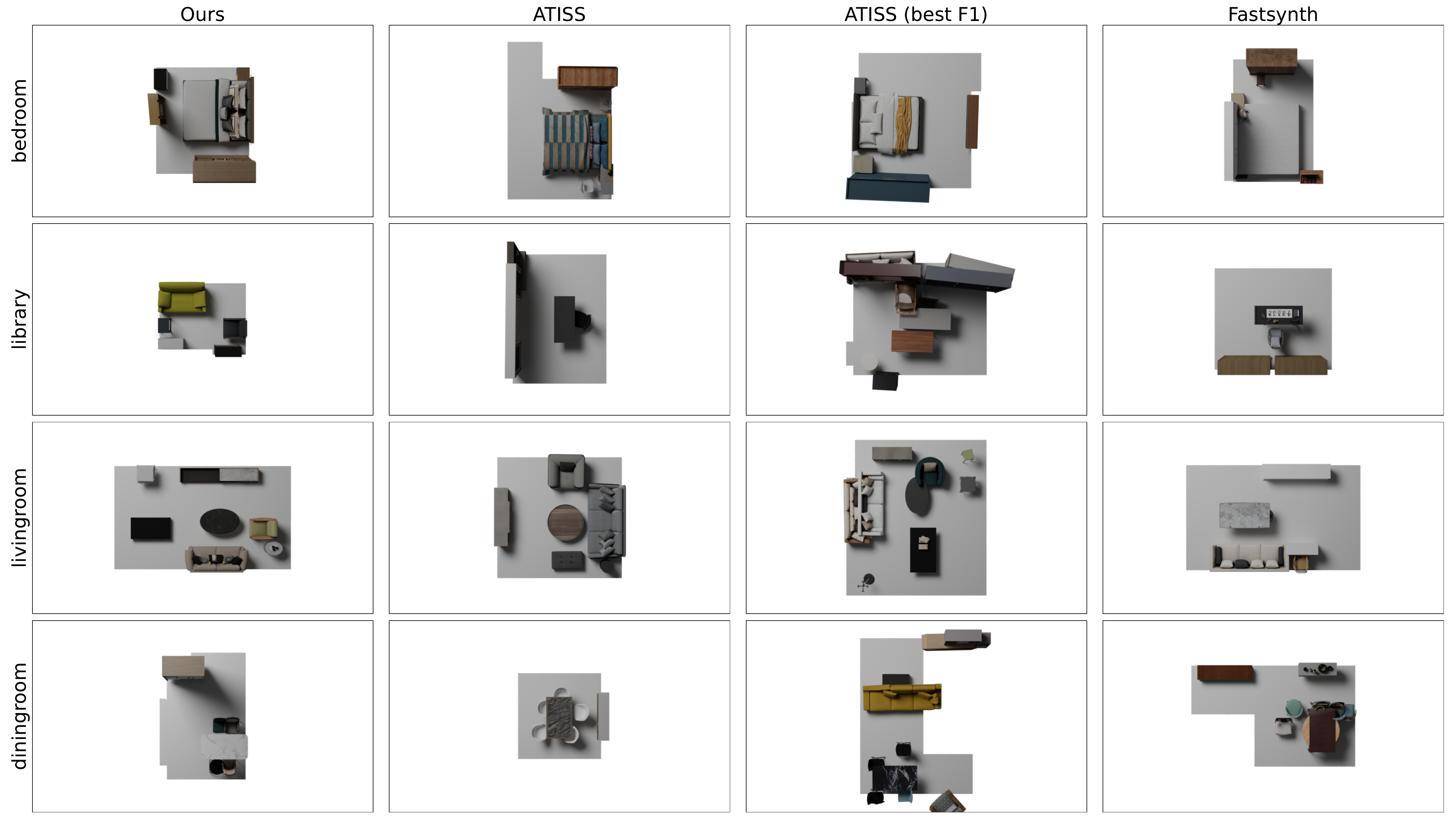}

   \caption{
   More examples of scenes generated by our and baseline methods
   }
   \label{fig:scene_diagram_2}
\end{figure*}

Figures \ref{fig:scene_diagram_1} and \ref{fig:scene_diagram_2} show more examples of scenes generated by our method and baseline methods

\section{Further Details on Constraints}
\label{sec:further_details_on_constraints}

\begin{figure*}[t]
  \centering
  \includegraphics[width=\linewidth]{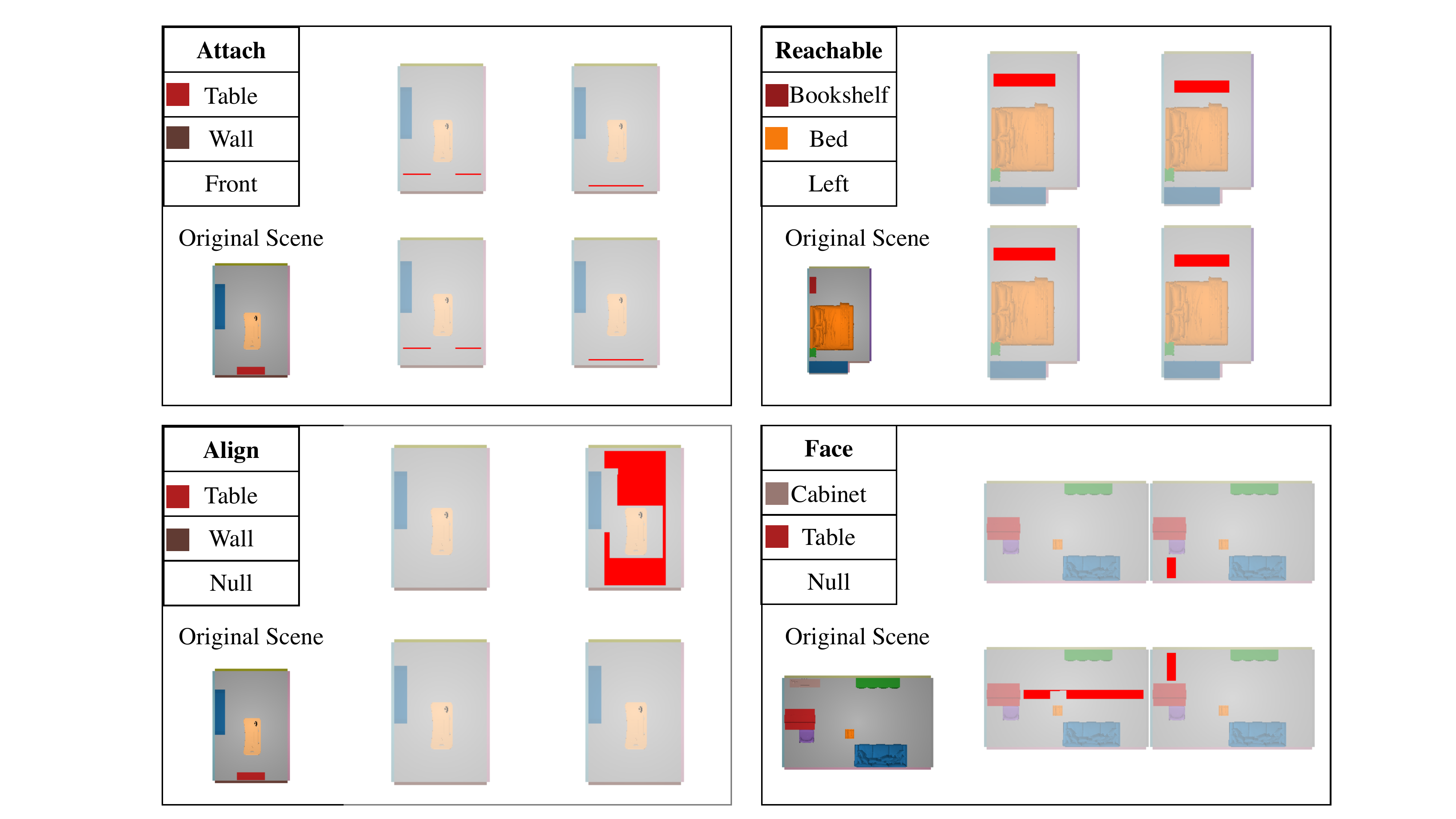}

   \caption{Constraint Examples: Shown are examples of constraints, their input scene and object, and their executed masks. Objects are colored by how they appear in the scene visualization. The original scene shows where the query object was originally placed. There are four masks for each constraint. Each mask represents a possible orientation of the query object. For example the align constraint contains placement options for only one orientation (the orientation of the reference wall object). The other constraints contain placement options for all possible orientations, but those locations are constrained based on the input arguments. 
   }
   \label{fig:example_constraints}
\end{figure*}

We assume the input objects to each constraints have the following assumed properties. Objects in each category are aligned in a canonical coordinate frame. They must also come labeled with whether they are meant for holding humans (i.e.  beds and chairs). Objects and their original scenes must also come with semantically meaningful sizes, scales, and distances since geometric heuristics described for each constraint are based on physically meaningful quantities such as the average reaching distance. Objects and scenes which satisfy this criteria come from preprocessing the 3DFRONT~\cite{fu20213d} dataset. Visualizations of each constraint is shown in Figure \ref{fig:example_constraints}. 

\section{Initial Program Extraction}
\label{sec:initial_program_extraction}
As a pre-processing step, scenes with major inter-object bounding box collisions are removed from the dataset. These scenes can produce errors in the scene extraction process. 

For every query object we first consider all the objects within attachment distance (15 cm). If a reference object is within attachment distance and also faces the same direction as the query object, an alignment constraint is also applied. Otherwise, if the two objects face each other, a face constraint is applied. If the query object is meant to hold humans such as a bed or a chair, the same process is applied for all objects within reaching distance (15 - 60 cm). 

This process is very sensitive to hyper parameters and can often fail to produce valid programs. In the case where a null program is extracted (a program that produces no object placements), we search through its children and if a subtree produces a program which contains the original placement, it is accepted. It is also often the case that too many constraints are applied and there are "extraneous" constraints, or constraints which when applied do not change the final output. These constraints are also removed. 

\section{Object Placement Classifier Architecture}
We perform experiments on two architectures of object placement classifiers. We test a CNN based classifier that uses a top-down image representation ~\cite{Ritchie2018FastAF} with an additional input channel denoting the query object in the scene. We also test a transformer-based classifier that uses the object encoder described in Section \ref{sec:generating_programs} with an additional learned vector embedding added to the query object vector encoding. While our CNN based classifier reports higher precision (accepts fewer invalid programs), and results in better downstream quantitative metrics, we report results using the transformer classifier due to constraints on computation. The transformer model is computationally cheaper than the CNN because it does not require rasterization of the top-down view of the scene. 

\section{Scene annotation software}
\label{sec:scene_annotation_software}

\begin{figure}[t]
    \centering
    \includegraphics[width=\linewidth]{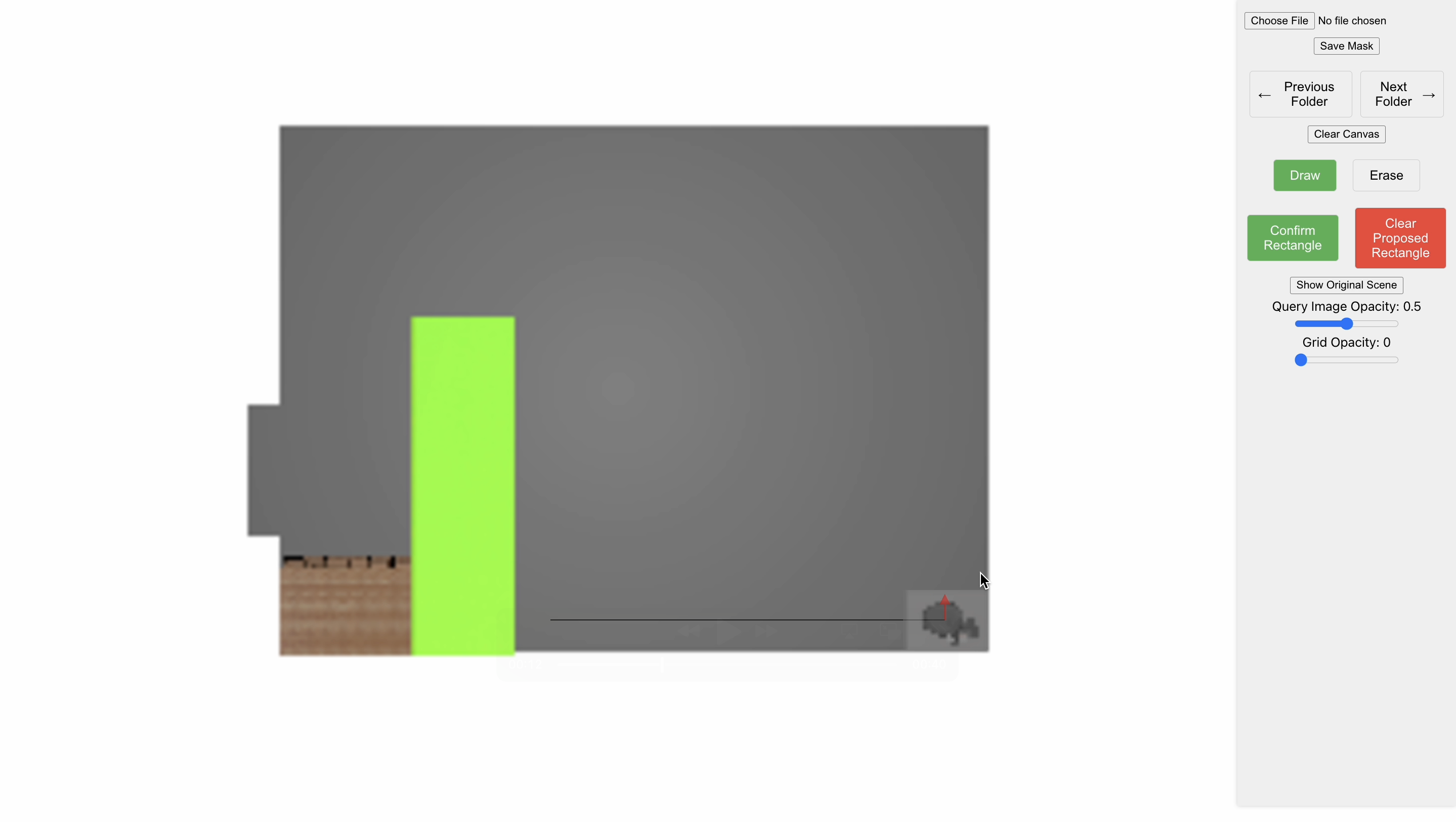}
    \caption{A screen shot of our annotation software. On the left shows a partial scene and the object hovers on the mouse pointer. Users can draw rectangles to visualize possible placements, and then confirm them once confident in their drawing. }
    \label{fig:scene-annotation}
\end{figure}

We built a browser based scene annotation software to facilitate the annotation of partial scenes, objects, and where they could go. We recruited 15 university students and young working professionals to participate in the annotation. Participants were given partial scene and object and asked to mark all the possible places that object can go. No time limit was enforced, but users spent an average of 18 seconds on each partial scene and object, or 30 minutes in total for 100 scenes.

The scene annotation software, built in react, allows users to draw rectangles denoting possible centroid locations of the object for a given orientation. Users can visualize what these proposed placements look like in the scene before confirming these placements. A screenshot of this software are shown in Figure \ref{fig:scene-annotation}

\section{Baselines Details}
The retraining of both baseline methods \cite{Paschalidou2021NEURIPS, Ritchie2018FastAF} on 3D-FRONT \cite{fu20213d} differ slightly from their original training setting. Both baselines are trained on the same scenes our method trains on and both without object ordering. 

ATISS's \cite{Paschalidou2021NEURIPS} data preprocessing parses living rooms and dining rooms with a maximum side length of 13.2 meters, and bedrooms and libraries with a maximum side length of 6.2. Our data processing only parses scenes with maximum side of 6.2 for all room types. During our data preprocessing, scenes with significant inter-object penetration are removed from the dataset as they introduce errors to the initial program extraction process. This also reduces the total number of scenes. in our data split.  

Fastsynth \cite{Ritchie2018FastAF} was not originally trained on 3D-FRONT. It was also trained with object ordering. We retrained Fastsynth on the same data splits of 3D-FRONT as our method, and also without object ordering for a fair comparison with ATISS.

\section{Edge Attention mechanism}

\begin{figure*}[ht]
    \centering
    \includegraphics[width=0.5\linewidth]{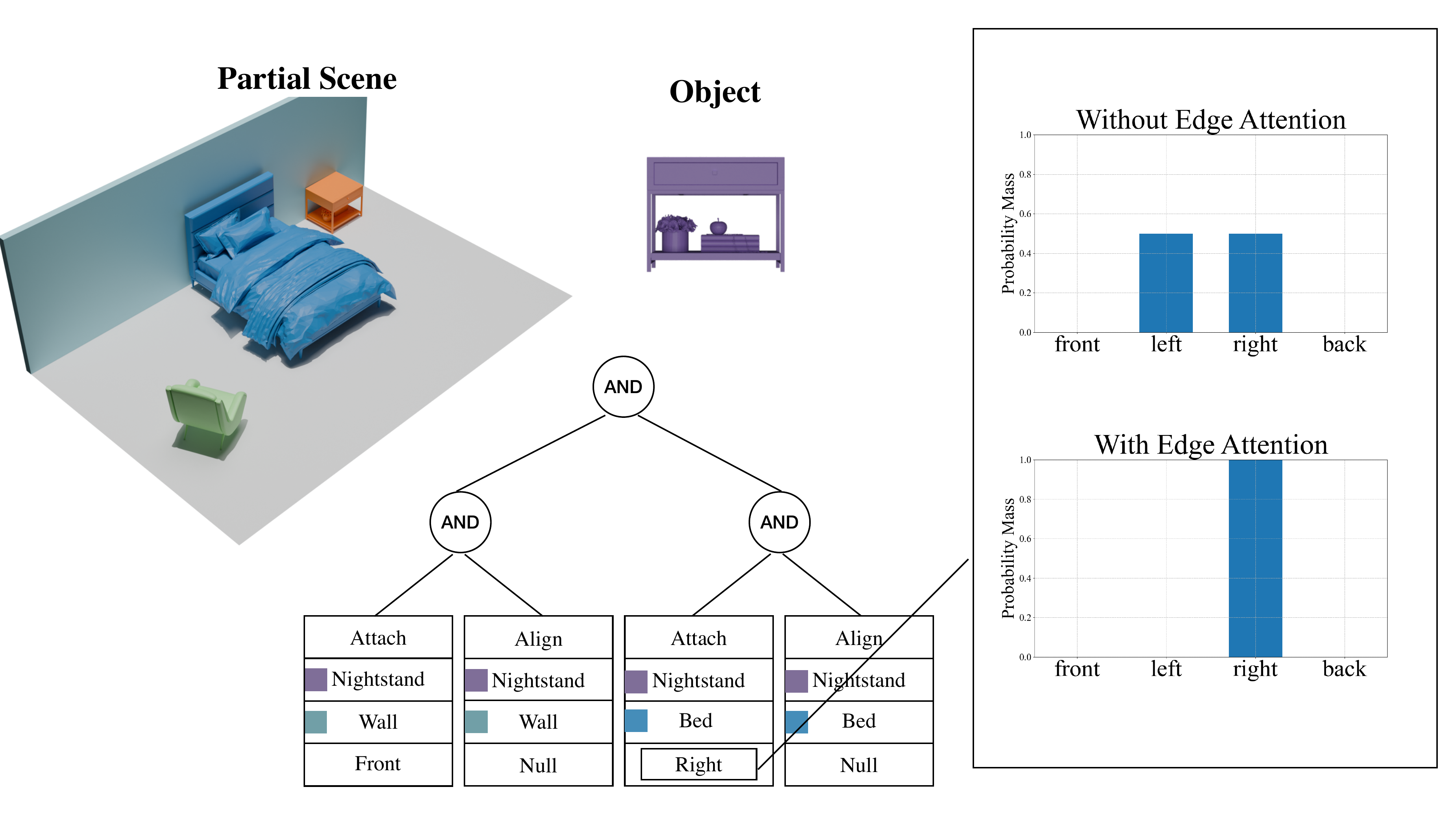}
    \caption{In this example, we want to place a nightstand in a room that already contains a bed and a nightstand on the left side of it. The program which specifies this placement should always predict placement on the right side of the bed. Without edge attention the program will incorrectly place the nightstand on the left side 50\% of the time. 
    With edge attention attention our model correctly attends to the spatial relationships between objects in the scene and predicts placement on the right side every time. 
    }
    \label{fig:edge_attention_example}
\end{figure*}

We describe the edge attention mechanism that augments our base transformer model. 

Consider the setting in Figure \ref{fig:edge_attention_example}. The partial scene contains a single bed and nightstand to the right of it. The query object is also a nightstand. Given this setting our generative model should predict a program that places the nightstand on the left of the bed 100 \% of the time. Instead, the logits corresponding to which side of the bed the program will place the nightstand is split 50-50 left and right. This is likely due to nightstands appearing on either side of the bed with equal frequency.  We find this is empirically true on both a model trained on real scene data and a toy setting containing just beds and nightstands. 

This experiment demonstrates that ordinary attention does not correctly account for spatial relationships between objects in our model. One interpretation of a transformer is that it is an edgeless graph neural network. Edge values denoting which side an object is on in relation to another object should provide the necessary information the network needs to correctly reason over the spatial relationships of objects in the room. As such, we augment the attention mechanism in our transformer model to introduce inter object relationships to the input signal.

In ordinary attention key, query, and value vectors are computed from linear projections of the input. The key and query vectors compute the attention weights used for a final weighted sum of the Value vectors. The output of regular self-attention $Z$ is defined as

\begin{equation}
   Z = \text{Softmax}(\frac{W_qX (W_kX)^T}{\sqrt{d_k}}) W_vX 
\end{equation}

Our edge attention mechanism adds inter object information to this computation. We extract directional relationships between objects and encode them into a matrix of edge values $E \in \mathbb{R}^{t x t x d}$.  Encoding directional relationships means that each object must receive its own matrix of edge values. We allow the original embedding vector inform which edges should receive more weight. The category, size, and location information encoded in the original embedding vector of an object should inform which other objects it pays attention to. For example, a bed should pay more attention to its spatial relationship with a nightstand than with a chair.

For each object, denoted by its index $i$, we compute key and value vectors $K’$ and $V’$ with respect to it. The attention weights are computed as $Q K^T + Q K’^T$, where the $QK’^T$ term acts as a correction weight to the original $QK^T$ attention weights. The weighted sum of $V’$ using these attention weights is added to the normal output of attention for the object. Given edge values $E_i \in \mathbb{R}^{t x d}$, our edge attention mechanism adds inter object information to each output of the original attention mechanism $Z_i$.

\begin{equation}
    Z_i = Z_i + \text{Softmax}(\frac{W_qX (W_kX)^T + W_qX (W_{ek}E_i)^T}{\sqrt{d_k}}) W_{ev}E_i
\end{equation}

\begin{figure*}[t]
  \centering
  \includegraphics[width=\linewidth]{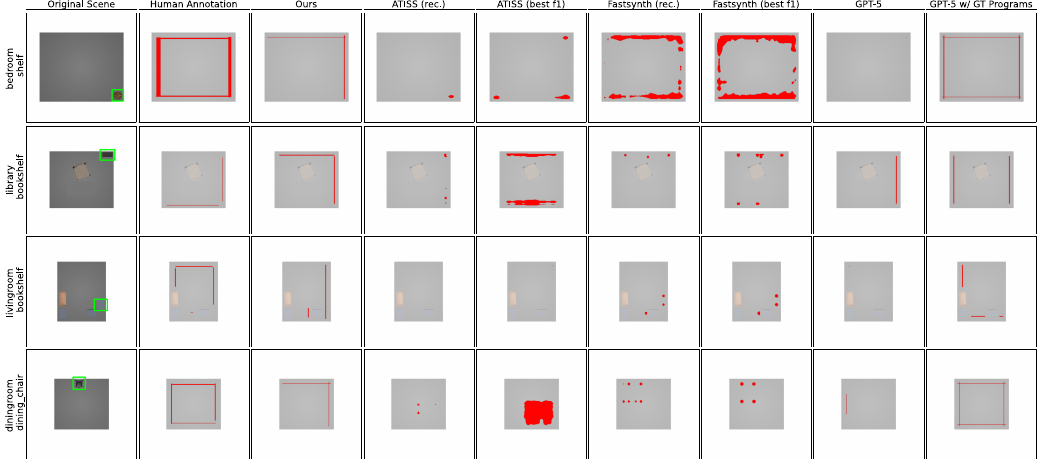}

   \caption{
   More examples of annotated masks alongside the location distributions predicted by our method and other methods. 
   }
\end{figure*}

\begin{figure*}[t]
  \centering
  \includegraphics[width=\linewidth]{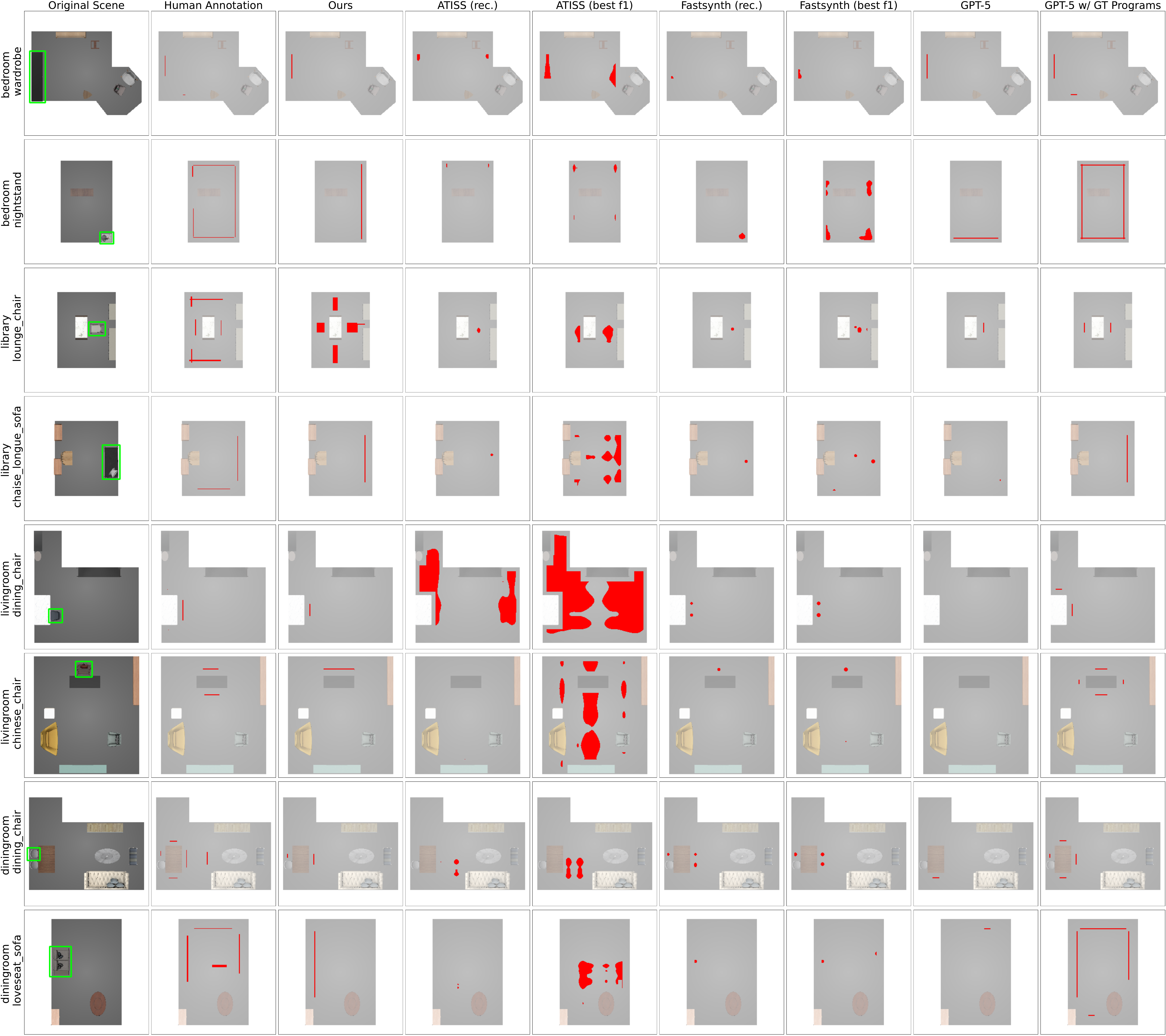}

   \caption{
   More examples of annotated masks alongside the location distributions predicted by our method and other methods. 
   }
\end{figure*}

\begin{figure*}[t]
  \centering
  \includegraphics[width=\linewidth]{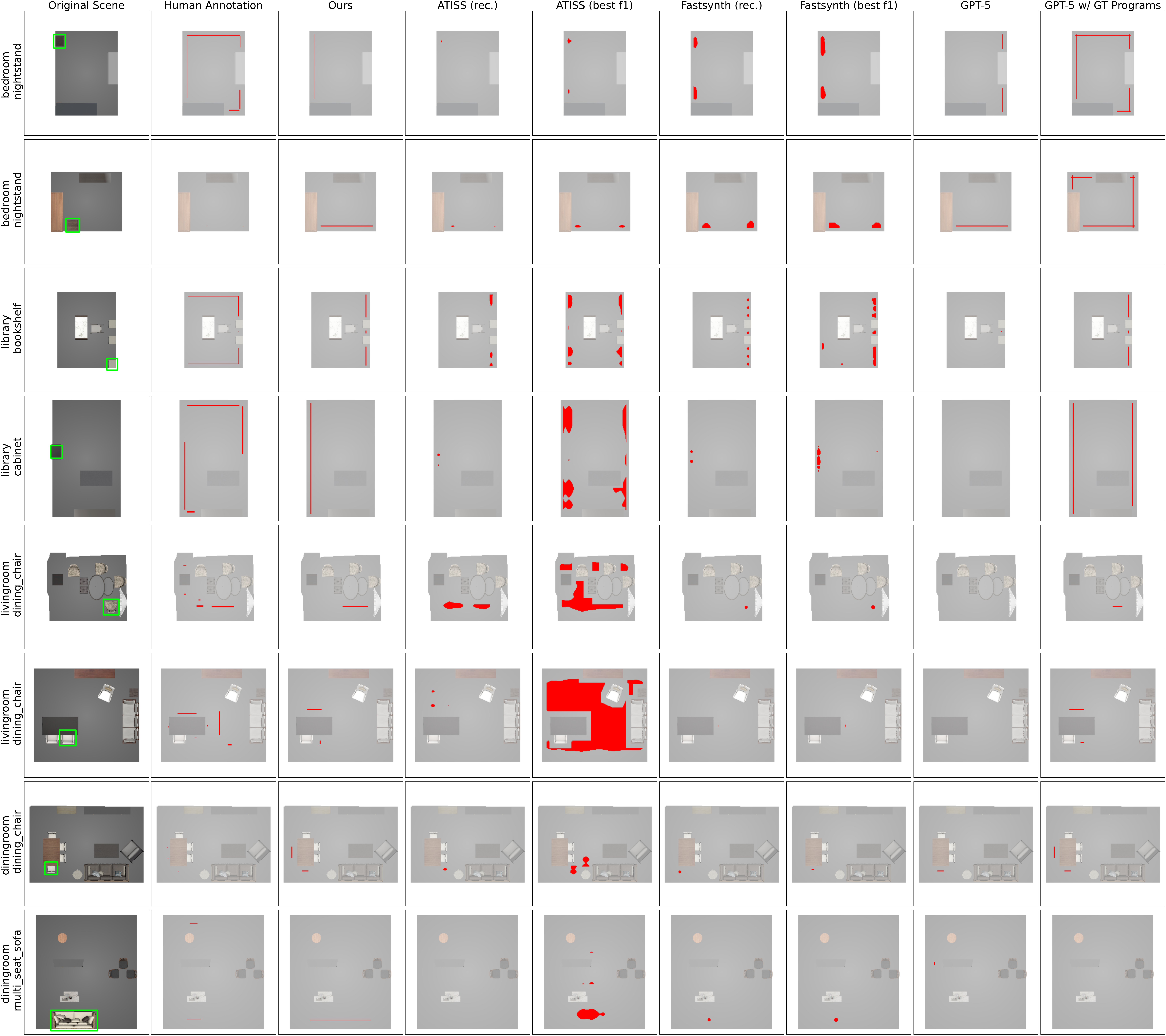}

   \caption{
   More examples of annotated masks alongside the location distributions predicted by our method and other methods. 
   }
\end{figure*}

\begin{figure*}[t]
  \centering
  \includegraphics[width=\linewidth]{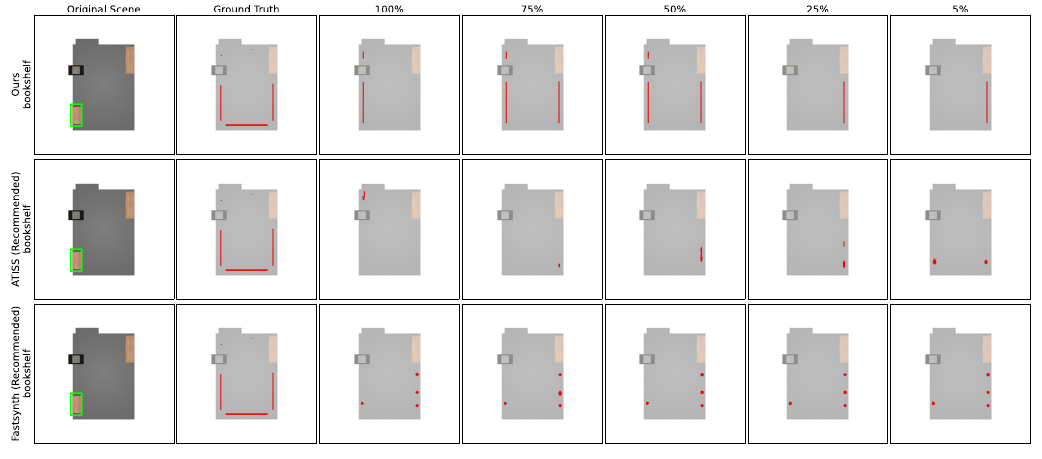}

   \caption{
   More examples of how less training data affects the predicted location distributions of our method and baselines
   }
\end{figure*}

